\documentclass[twocolumn,superscriptaddress,amssymb,amsmath,nobibnotes,aps,prd,showpacs,nofootinbib]{revtex4}%
\pdfoutput=1

\usepackage{graphicx,float}
\usepackage{epsf}
\usepackage{bm}
\usepackage{amsmath,hyperref}
\usepackage{amsfonts}
\usepackage{amssymb}
\usepackage{epstopdf}
\usepackage{natbib}%
\usepackage{rotating}
\usepackage{pdflscape}
\usepackage{adjustbox}

\usepackage[toc,page]{appendix}
\providecommand{\U}[1]{\protect\rule{.1in}{.1in}}
\newcommand{\be}{\begin{equation}}
\newcommand{\ee}{\end{equation}}

\newcommand{\mincir}{\raise
-3.truept\hbox{\rlap{\hbox{$\sim$}}\raise4.truept\hbox{$<$}\ }}
\newcommand{\magcir}{\raise
-3.truept\hbox{\rlap{\hbox{$\sim$}}\raise4.truept\hbox{$>$}\ }}

\usepackage{color}
\definecolor{darkgreen}{rgb}{0., 0.65, 0.1}
\begin{document}

\title{\textcolor{blue}{\small{$\quad$\hspace{2.5cm}PHYSICAL REVIEW D \textbf{96} (2017) 103511 }}\newline\newline Echo of interactions in the dark sector}

\author{Suresh Kumar}
\email{suresh.kumar@pilani.bits-pilani.ac.in}
\affiliation{Department of Mathematics, BITS Pilani, Pilani Campus, Rajasthan-333031,India}
 
\author{Rafael C. Nunes}
\email{rafadcnunes@gmail.com}
\affiliation{Departamento de F\'isica, Universidade Federal de Juiz de Fora, 36036-330,
Juiz de Fora, MG, Brazil}
 

%%%%%%%%%%%%%%
\begin{abstract}
We investigate the observational constraints on an interacting vacuum energy scenario with two different neutrino schemes (with and without a sterile neutrino)
using the most recent data from CMB temperature and polarization anisotropy, 
baryon acoustic oscillations (BAO), type Ia supernovae from JLA sample and structure growth inferred from cluster counts. 
We find that inclusion of the galaxy clusters data with the minimal data combination CMB + BAO + JLA suggests an interaction in the dark sector, 
implying the decay of dark matter particles into dark energy, since the constraints obtained by including the galaxy clusters data yield a
non-null and negative coupling parameter between the dark components at 99\% confidence level. 
We deduce that the current tensions on the parameters $H_0$ and $\sigma_8$ can be alleviated within the framework of the interacting as well as non-interacting vacuum energy models with sterile neutrinos.

\end{abstract}

\maketitle

\section{Introduction}
\label{sec:intro}

Since the discovery that our Universe is in a current stage of accelerated expansion \cite{int1,int2}, 
the inclusion of an exotic energy component with negative pressure, dubbed as dark energy (DE), in the energy budget of the Universe is necessary to explain the observations.
A cosmological model with cold dark matter (CDM) and DE mimicked by a positive cosmological constant $\Lambda$, the so-called $\Lambda$CDM model,
is considered as the standard cosmological model for its ability to fit the observational data with great precision.
A large amount of data, pouring in from different physical origins, have been measured with good precision 
to unveil the dynamics/mysteries of the Universe in the last decade or so. Some data sets are found in tension with others within the framework of the $\Lambda$CDM model (see\cite{T1,T2,T3,T4,T5}), which could be an indication of new 
physics beyond $\Lambda$CDM model or may be due to some systematic effects on the data. Also, the cosmological constant 
suffers from some theoretical problems \cite{int3,int4a,int4b,int5}, which motivated the researchers to 
propose new models of DE, or alternative mechanisms that can explain the observational data without the need for DE, 
such as modified gravity models. Cosmological models, where dark matter (DM) and DE interact throughout the evolution history 
of the Universe, have received considerable attention of the researchers in the recent past in order to 
solve/assuage the problem of the cosmic coincidence as well as the problem 
of the cosmological constant (for the case where DM interacts with vacuum energy). 
See \cite{DE_DM_1,DE_DM_2} for general review. 
The late-time interaction in the dark sector has recently been investigated in various studies  
\cite{Salvatelli,Sola1,Sola2,Saulo,Richarte,Valiviita,Elisa,Murgia:2016ccp,N1,N2,N3,N4,N5} in different contexts. 

On the other hand, physical experiments from solar, atmospheric, reactor and accelerator neutrino beams have proved 
the existence of neutrino oscillations (see \cite{Garcia} and references therein for review and \cite{Garcia2, Capozzi} for
current experimental status). The phenomenon of neutrino oscillations implies directly a non-zero neutrino mass, 
once the neutrino oscillations depend only on the mass difference between the three species predicted by 
the standard model of particle physics. Non-standard neutrinos, like the sterile massive neutrinos, 
are predicted in a natural extension of the standard model of elementary particles. 
Sterile neutrinos at the eV mass scale, can explain the anomalies found in some short-baseline neutrino 
oscillation experiments from the reactor antineutrino anomaly, the Gallium neutrino anomaly, 
and LSND experiments (see \cite{Palazzo,Abazajian,Gariazzo} for review). These experiments require 
the existence of light sterile neutrinos.

Massive neutrinos play an important role on cosmological scales, with direct implications on Cosmic Microwave Background (CMB) temperature  
and polarization anisotropy, large scale structure, nuclear abundances produced by big bang nucleosynthesis, 
and in some other cosmological sources (see \cite{Dolgov,Lesgourgues,Lesgourgues2} for review, Planck collaboration paper \cite{Planck2015}, 
and forecast from CORE space mission \cite{CORE} for observational constraints). Recently, cosmological models with massive neutrinos have been 
studied to reconcile the tension between local and global determinations of the Hubble constant 
\cite{Valentino,Bernal,Wyman,Riess,Tram}. The presence of massive neutrinos in cosmological models is useful 
to investigate the accuracy and robustness of a given scenario. The cosmological effects of light sterile neutrinos are 
investigated in different contexts \cite{m_esteril1,m_esteril2,m_esteril3,m_esteril4,m_esteril5,m_esteril6}

In this paper, we are motivated to investigate a coupled DE 
scenario, where vacuum energy interacts with DM, by considering neutrinos scheme with sum of the active neutrino masses
($\sum m_\nu$) and the effective number of relativistic degrees of freedom ($N_{\rm eff}$) as free quantities. In addition, we consider the models with
one non-standard neutrino (sterile neutrino) beyond the three species predicted by the standard model, 
the so-called (3 + 1) models. We study observational constraints on the models by using the most recent data
from CMB temperature and polarization anisotropy, baryon acoustic
oscillations, type Ia supernovae from JLA sample, and galaxy clusters measurements. 
Our main result is that the coupling parameter of the dark sector is observed to be non-zero up to $99\%$ confidence level (CL). 

Rest of the paper is structured as follows: The interacting vacuum energy scenario is introduced in the following Section II while 
different models of neutrinos with the interacting as well as non-interacting vacuum energy scenario are described in Section III. 
The observational constraints from various recent data sets on the models under consideration are obtained and discussed 
in detail in Section IV. The conclusions the our investigation are summarized in the final section.

\section{Interaction in the dark sector}
\label{sec:model}
 
We consider a spatially-flat Friedmann-Robertson-Walker Universe, with the Friedmann equation given by

\begin{align}
\label{friedmann}
3 H^2 = 8 \pi G (\rho_{\gamma} + \rho_{\nu} + \rho_{\rm b} + \rho_{\rm dm} +
\rho_{\rm de}  ),
\end{align}
where $H= \dot{a}/a$ is the Hubble parameter. Further $\rho_{\gamma}$, $\rho_{\nu}$, $\rho_{\rm b}$,
$\rho_{\rm dm}$, and $\rho_{\rm de}$ stand for the energy densities of photons, massive neutrinos, baryons, cold DM and DE, 
respectively.

We assume that DM and DE interact via the coupling function $Q$ given by
\begin{align}
\label{Q_equation}
\dot{\rho}_{\rm dm} + 3H\rho_{\rm dm} = -\dot{\rho}_{\rm de} -
3H \rho_{\rm de} (1 + w_{\rm de}) = Q.
\end{align}
For $Q > 0$, 
the energy flow takes place from DE to DM, otherwise the energy flows from DM to DE.
The parameter $w_{\rm de}$ is the equation of state of the DE and in general can be 
constant or variable. In order to avoid many free parameters, we consider the simplest case 
$w_{\rm de} = -1$, i.e., an interacting vacuum energy scenario. 
In what follows, we refer this model to as IVCDM model. For the sake of comparison, we shall also consider the non-interacting vacuum energy case ($\delta=0$), i.e., the $\Lambda$CDM model.

Recent studies of interacting vacuum cosmologies have
focused on models with general interaction term given by $Q = \delta H \rho_{\rm dm}$ or $Q = \delta H \rho_{\rm de}$, where $\delta$ 
is a constant and dimensionless parameter that characterizes the interaction between DM and DE. 
Both classes have been investigated from different perspectives 
\cite{coupled01,coupled02,coupled03,coupled04,coupled05,coupled06,coupled07,coupled08,coupled09,coupled10,coupled11,coupled12}.
For instance, in\cite{delta_cdm3}, we investigated the interaction in dark sector given by the coupling 
function $Q = \delta H \rho_{\rm dm}$.  We noted that this coupling function may suffer 
from instability in the dark sector perturbations at early times 
(the curvature perturbations may blow up on super Hubble scales), and to avoid such problems 
we should have $\delta < 10^{-2}$. On the other hand, it is known that $Q = \delta H \rho_{\rm de}$ does not present these 
problems \cite{delta_cdm4}, and that the coupling parameter $\delta$ can admit much larger values. 
Once we consider data from perturbative origin in our analysis, the choice $Q = \delta H \rho_{\rm de}$  
offers a greater range of security/freedom over the coupling parameter, 
and thus allows this parameter to be constrained within a larger prior range. For instance, the authors in \cite{Salvatelli} 
used this coupling function to show that a general late-time interaction between cold DM and vacuum energy is
favored by current cosmological datasets. Therefore, in the present study, we use the coupling function given 
by $Q = \delta H \rho_{\rm de}$. Once the coupling function $Q$ has been defined, 
the background evolution of the dark components can be obtained from eq. (\ref{Q_equation}).
Here, we adopt the synchronous gauge for the evolution of the scalar mode perturbations of the dark components as 
in \cite{delta_cdm,delta_cdm2}, or more explicitly the eqs. (6) - (9) of \cite{delta_cdm3}. The baryons, photons and massive neutrinos are 
conserved independently, and the perturbation equations follow the standard evolution as described in 
\cite{Ma_Bertschinger}. 

\section{The models}
\label{model}

On the basis of neutrinos, we consider the following
models in our study.\\

$\checkmark$ We consider the interacting and non-interacting vacuum energy scenarios possessing neutrinos scheme with sum of the active neutrino masses
($\sum m_\nu$) and the effective number of relativistic degrees of freedom ($N_{\rm eff}$) as free quantities. In our results, we shall denote these models as 
$\nu$IVCDM and $\nu\Lambda$CDM, respectively. 
\\

% We intend to investigate possible correlation of $N_{\rm eff}$ with the coupling parameter $\delta$, and  other parameters of interest. 
% The base parameters set for the $\nu$IVCDM  model is
% 
% \begin{eqnarray*}
% \label{P1}
% P = \{100\omega_{\rm b}, \, \omega_{\rm dm}, \, 100\theta_{s}, \, \ln10^{10}A_{s}, \, \nonumber \\ 
% n_s, \, \tau_{\rm reio}, \, N_{\rm eff}, \, \sum m_{\nu}, \, \delta \}.
% \end{eqnarray*}

$\checkmark$ Following the Planck collaboration, we fix the mass ordering of the active neutrinos 
to the normal hierarchy with the minimum masses allowed by oscillation experiments, i.e., $\sum m_{\nu} = 0.06$ eV. As generalization, let us take the mass order, $m_1 = m_2 = 0$ eV 
and $m_3 = 0.06$ eV. Therefore, the presence of excess mass can be considered from a single additional mass 
state $m_4$, which can be related to the sterile neutrino mass, $m_{{\nu}_s}$. Taking this into account, 
the mass splitting with relation to the sterile neutrino can be written as $\Delta m_{41} = m_4$.
In order to constrain the sterile neutrino mass, the effective number of neutrino species is $N_{\rm eff} = 4.046$, i.e., 
we consider one species of sterile neutrinos (fully thermalized) beyond the three neutrino species of the standard model, 
the so-called (3+1) models. We consider the sterile neutrino case with the interacting and non-interacting vacuum energy scenarios, 
and  denote these models as $\nu_s$IVCDM and $\nu_s\Lambda$CDM, respectively.

% The base parameters set for the $\nu_s$IVCDM  model is
% 
% \begin{eqnarray*}
% \label{P1}
% P = \{100\omega_{\rm b}, \, \omega_{\rm dm}, \, 100\theta_{s}, \, \ln10^{10}A_{s}, \, \nonumber \\ 
% n_s, \, \tau_{\rm reio}, \, m_{\nu_s}, \, \delta \}.
% \end{eqnarray*}

\section{Observational Constraints}

In order to constrain the models under consideration, we use the following data sets. 
\\

\textbf{CMB}: The cosmic microwave background (CMB) data from Planck 2015 comprised of the likelihoods 
at multipoles $l \geq 30$ using TT, TE and EE power spectra and the low-multipole polarization likelihood at $l \leq 29$. The combination of these data is referred to as Planck TT, TE,
EE + lowP in \cite{Planck2015}. We also include Planck 2015 CMB lensing 
power spectrum likelihood \cite{CMB}. 

\textbf{BAO}: The baryon acoustic oscillations (BAO) measurements from the  Six  Degree  Field  Galaxy  Survey  (6dF) \cite{bao1}, 
the  Main  Galaxy  Sample  of  Data  Release 7  of  Sloan  Digital  Sky  Survey  (SDSS-MGS) \cite{bao2}, 
the  LOWZ  and  CMASS  galaxy  samples  of  the Baryon  Oscillation  Spectroscopic  Survey  (BOSS-LOWZ  and  BOSS-CMASS,  
respectively) \cite{bao3},  and the distribution of the LymanForest in BOSS (BOSS-Ly) \cite{bao4}. 
These data points are summarized in table I of \cite{coupled04}.

\textbf{JLA}: The latest ``joint light curves" (JLA) sample \cite{snia3}, comprised of 740 type Ia supernovae
in the redshift range $z \in [0.01, 1.30]$. 

% \textbf{CC+$H_0$}: The cosmic chronometers (CC) data set comprising of 30 measurements 
% spanned in the redshift range $0 < z < 2$,  recently compiled in \cite{cc}. We also use the recently measured new local
% value of Hubble constant given by $H_0=73.24 \pm 1.74$  km s${}^{-1}$ Mpc${}^{-1}$ as reported in \cite{Riess}. 

\textbf{GC}: The measurements from the abundance of galaxy clusters (GC) offer a powerful 
probe of the growth of cosmic structures. The cosmological information enclosed in the cluster abundance is efficiently parameterized
by $S_8 = \sigma_8(\Omega_m/\alpha)^{\beta}$, where $\sigma_8$ is the linear amplitude of the fluctuations on 8 Mpc/h
scale and $\alpha$, $\beta$ are adopted/fit parameters in each survey analysis.
We use the measurements from the Sunyaev-Zeldovich effect cluster mass function measured by Planck (SZ) 
\cite{cg1} ($\alpha = 0.27$ and $\beta = 0.30$) and 
the constraints derived by weak gravitational lensing data from Canada-France-Hawaii Telescope Lensing Survey (CFHTLenS) 
\cite{cg2} ($\alpha =0.27 $ and $\beta = 0.46$).

We modified the publicly available CLASS \cite{class} and Monte Python \cite{monte} codes for the models considered in the present work. 
We used Metropolis Hastings algorithm with uniform priors on the model parameters to obtain correlated Markov Chain Monte Carlo samples by 
considering four combinations of data sets: CMB + BAO + JLA, CMB + BAO + JLA + CFHTLenS, CMB + BAO + JLA + SZ, and CMB + BAO + JLA + CFHTLenS + SZ.
All parameters in our Monte Carlo Markov Chains converge according to the Gelman-Rubin criteria \cite{Gelman}.
The resulting samples are then analysed by using the GetDist Python package \cite{antonygetdist}. Note that we have chosen the 
minimal data set CMB+BAO+JLA because adding BAO + JLA data to CMB does not shift the regions of probability much, but this combination constrains 
the matter density very well, and reduces the error-bars on the parameters. 
The other data set combinations are considered to investigate how the constraints on various 
free parameters and derived parameters are affected by the inclusion of GC data.
The final statistical results are summarized in Table \ref{tab1}.

\begin{figure*}
\includegraphics[width=12.0cm]{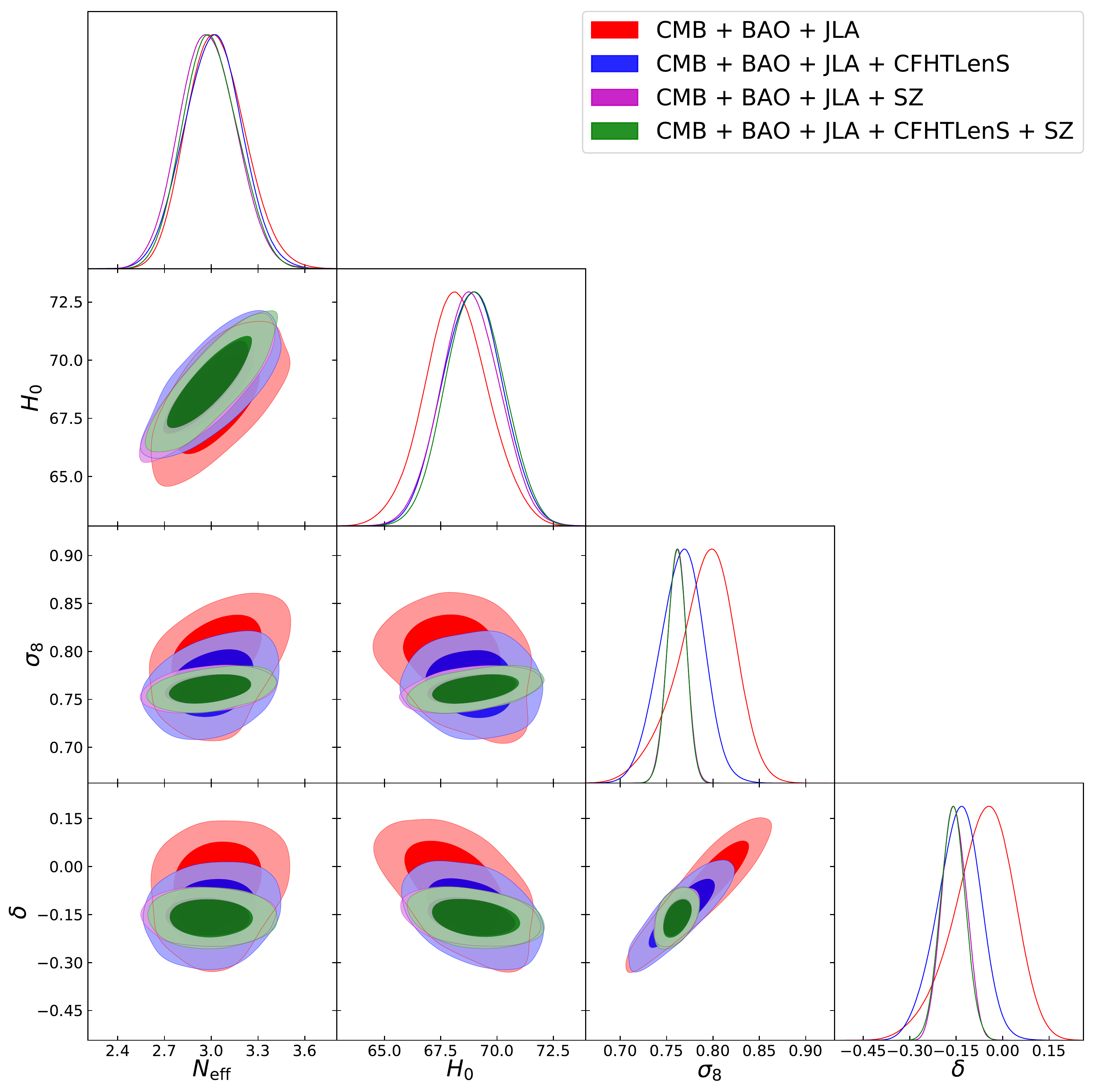}
\caption{\label{fig1} {\it{One-dimensional marginalized distribution, and 68\% and 95\% CL regions 
for some selected parameters of the $\nu$IVCDM model.}}}
\end{figure*}

\begin{figure*}
\includegraphics[width=12.0cm]{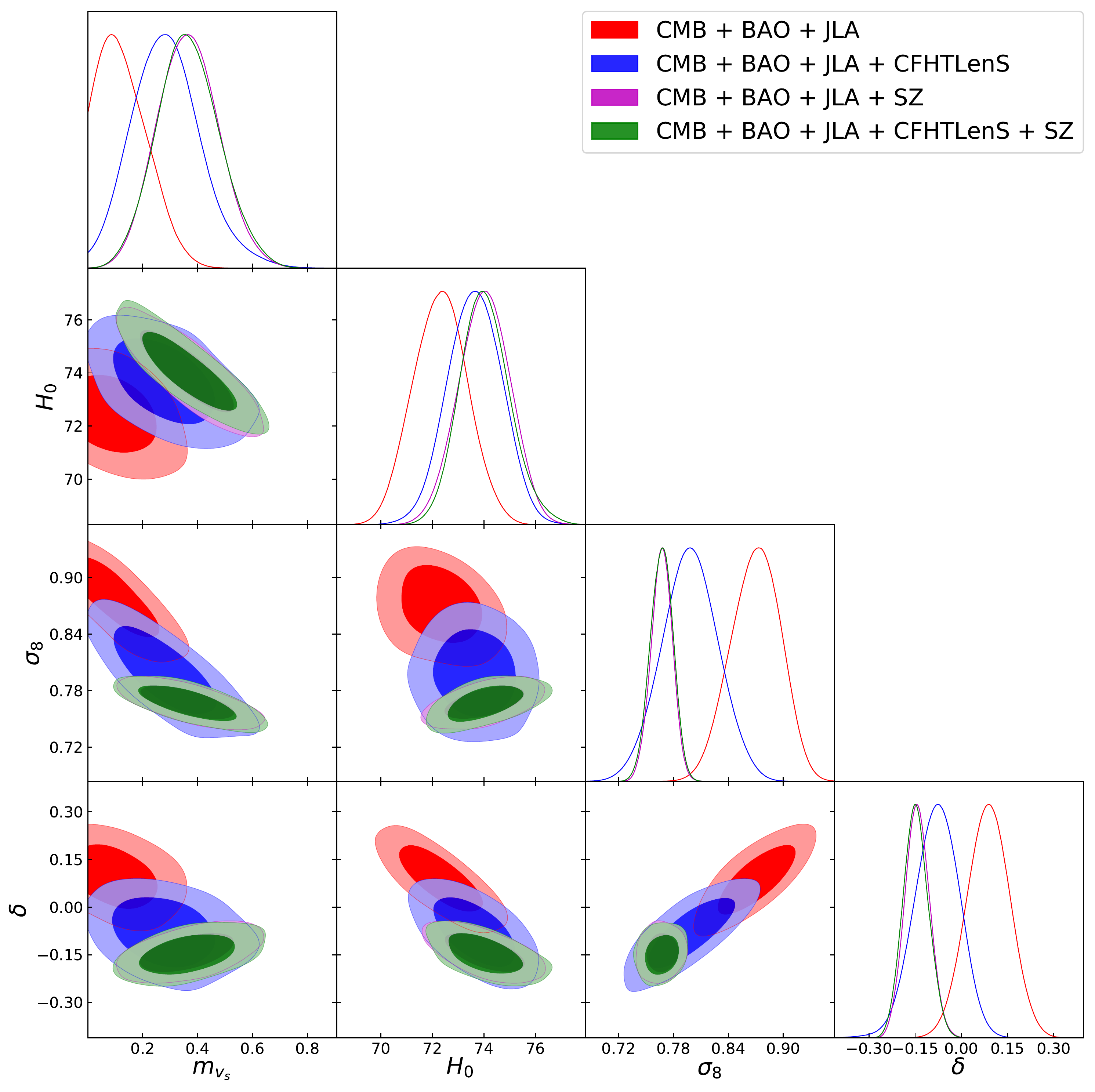}
\caption{\label{fig2} {\it{One-dimensional marginalized distribution, and 68\% and 95\% CL regions 
for some selected parameters of the $\nu_s$IVCDM model.}}}
\end{figure*}

%#################################################################################################################################
\begin{table*}[!htbp]
\caption{\label{tab1} Constraints on free parameters and two derived parameters $H_0$ and $\sigma_8$
of the $\nu$IVCDM, $\nu \Lambda$CDM, $\nu_s$IVCDM and $\nu_s \Lambda$CDM models. For each parameter, the four entries 
(mean values with 1$\sigma$ errors) are the constraints from the data combinations \textcolor{red}{CMB + BAO + JLA}, \textcolor{blue}{CMB + BAO + JLA + CFHTLenS}, \textcolor{magenta}{CMB + BAO + JLA + SZ} and  \textcolor{darkgreen}{CMB + BAO + JLA + CFHTLenS + SZ}, respectively. The parameter $H_0$ is in the units 
of km s${}^{-1}$ Mpc${}^{-1}$, while $\sum m_{\nu}$ and $m_{{\nu}_s}$ are in the units of eV. The $\chi^2_{\rm min}$ values of the fit are also displayed.}
     \begin{center}
\begin{tabular} { l  l lll}
\hline
   & $\nu$IVCDM & $\nu \Lambda$CDM       & $\nu_s$IVCDM & $\nu_s \Lambda$CDM \\
\hline
\hline
{$10^{2}\omega_{\rm b }$} &$\textcolor{red}{2.230\pm 0.021}$     & $\textcolor{red}{2.238\pm 0.019}$    &$\textcolor{red}{2.314\pm 0.014}$  &$\textcolor{red}{2.311\pm 0.016}$   \\
 &$\textcolor{blue}{2.226\pm 0.021}$     &$\textcolor{blue}{2.241\pm 0.020}$     &$\textcolor{blue}{2.308\pm 0.015}$  &$\textcolor{blue}{2.314\pm 0.015}$  \\
 &$\textcolor{magenta}{2.222\pm 0.020}$     &$\textcolor{magenta}{2.241\pm 0.022}$     &$\textcolor{magenta}{2.308\pm 0.014}$  & $\textcolor{magenta}{2.323\pm 0.014 }$ \\
 &$\textcolor{darkgreen}{2.224\pm 0.020}$     &$\textcolor{darkgreen}{2.241\pm 0.021}$     &$\textcolor{darkgreen}{2.307\pm 0.015}$  & $\textcolor{darkgreen}{2.326\pm 0.015}$ \\
                                                                                                
{$\omega_{\rm {cdm}}$}  &$\textcolor{red}{0.1182\pm 0.0028}$     & $\textcolor{red}{0.1188^{+0.0028}_{-0.0031}}$    &$\textcolor{red}{0.1322\pm 0.0013}$  & $\textcolor{red}{0.1326\pm 0.0015}$  \\
 &$\textcolor{blue}{0.1182\pm 0.0029}$     &$\textcolor{blue}{0.1181\pm 0.0029}$     &$\textcolor{blue}{0.1325\pm 0.0015}$  &$\textcolor{blue}{0.1318\pm 0.0014 }$ \\
 &$\textcolor{magenta}{0.1178\pm 0.0029}$     &$\textcolor{magenta}{0.1152\pm 0.0029}$     &$\textcolor{magenta}{0.1324\pm 0.0015}$  & $\textcolor{magenta}{0.1293\pm 0.0012}$  \\
 &$\textcolor{darkgreen}{0.1181\pm 0.0028}$     &$\textcolor{darkgreen}{0.1149\pm 0.0030}$     &$\textcolor{darkgreen}{0.1325\pm 0.0015}$  & $\textcolor{darkgreen}{0.1290\pm 0.0011}$  \\
                                                                                                
{$100\theta_{s }  $} &$\textcolor{red}{1.04200\pm 0.00049}$     &$\textcolor{red}{1.04185\pm 0.00049}$     &$\textcolor{red}{1.04005\pm 0.00029}$  &$\textcolor{red}{1.04000\pm 0.00030}$  \\
 &$\textcolor{blue}{1.04200\pm 0.00051}$     &$\textcolor{blue}{1.04190\pm 0.00048 }$     &$\textcolor{blue}{1.04005\pm 0.00029}$  &$\textcolor{blue}{1.04007\pm 0.00030}$  \\
 &$\textcolor{magenta}{1.04026\pm 0.00030  }$     & $\textcolor{magenta}{1.04219\pm 0.00053}$    &$\textcolor{magenta}{1.04002\pm 0.00029}$  & $\textcolor{magenta}{1.04032\pm 0.00028}$ \\
 &$\textcolor{darkgreen}{1.04200\pm 0.00049}$     & $\textcolor{darkgreen}{1.04221\pm 0.00053}$    &$\textcolor{darkgreen}{1.04002\pm 0.00032}$  & $\textcolor{darkgreen}{1.04032\pm 0.00028}$ \\
                                                                                              
{$\ln10^{10}A_{s }$}  &$\textcolor{red}{3.091^{+0.027}_{-0.030}}$     &$\textcolor{red}{3.091\pm 0.026}$     &$\textcolor{red}{3.149\pm 0.029}$  &$\textcolor{red}{3.158\pm 0.032}$  \\
 &$\textcolor{blue}{3.091\pm 0.028 }$     &$\textcolor{blue}{3.086\pm 0.028}$     &$\textcolor{blue}{3.171\pm 0.031}$  & $\textcolor{blue}{3.164\pm 0.031}$ \\
 &$\textcolor{magenta}{3.089^{+0.027}_{-0.031}  }$     & $\textcolor{magenta}{3.079^{+0.033}_{-0.039}}$    &$\textcolor{magenta}{3.184\pm 0.029}$  & $\textcolor{magenta}{3.190\pm 0.029}  $ \\
 &$\textcolor{darkgreen}{3.089^{+0.025}_{-0.030} }$     & $\textcolor{darkgreen}{3.076^{+0.034}_{-0.038}}$    &$\textcolor{darkgreen}{3.186\pm 0.028}$  & $\textcolor{darkgreen}{3.193\pm 0.028}  $ \\
                                                                                                
{$n_{s }         $}  &$\textcolor{red}{0.9667\pm 0.0082}$     & $\textcolor{red}{0.9703\pm 0.0080}$    &$\textcolor{red}{1.0050\pm 0.0043}$  &$\textcolor{red}{1.0040\pm 0.0043}$  \\
 &$\textcolor{blue}{0.9648\pm 0.0080}$     &$\textcolor{blue}{0.9703\pm 0.0076}$     &$\textcolor{blue}{1.0031\pm 0.0045}$  &$\textcolor{blue}{1.0049\pm 0.0043 }$  \\
 & $\textcolor{magenta}{0.9632\pm 0.0078}$    &$\textcolor{magenta}{0.9684\pm 0.0090 }$     &$\textcolor{magenta}{1.0023\pm 0.0043}$  &$\textcolor{magenta}{1.0061\pm 0.0041}$  \\
 & $\textcolor{darkgreen}{0.9640\pm 0.0074}$    &$\textcolor{darkgreen}{0.9680\pm 0.0087}$     &$\textcolor{darkgreen}{1.0021\pm 0.0043}$  &$\textcolor{darkgreen}{1.0070\pm 0.0041}$  \\
                                                                                                
{$\tau_{\rm reio }   $} &$\textcolor{red}{0.080\pm 0.014}$     &$\textcolor{red}{0.079\pm 0.013}$     &$\textcolor{red}{0.094\pm 0.016}$  &$\textcolor{red}{0.098\pm 0.017}$  \\
 &$\textcolor{blue}{0.081\pm 0.014  }$     &$\textcolor{blue}{0.078\pm 0.014 }$     &$\textcolor{blue}{0.105\pm 0.017}$  &$\textcolor{blue}{0.102\pm 0.017}$  \\
 &$\textcolor{magenta}{0.080\pm 0.014}$     &$\textcolor{magenta}{0.078^{+0.016}_{-0.018}}$     &$\textcolor{magenta}{0.113\pm 0.015}$  &$\textcolor{magenta}{0.118\pm 0.016}$  \\
 &$\textcolor{darkgreen}{0.079^{+0.013}_{-0.015}}$     &$\textcolor{darkgreen}{0.077^{+0.016}_{-0.018}}$     &$\textcolor{darkgreen}{0.113\pm 0.015}$  &$\textcolor{darkgreen}{0.120\pm 0.015}$  \\

{$N_{\rm eff }   $} &$\textcolor{red}{3.03^{+0.17}_{-0.20}}$    &$\textcolor{red}{3.10\pm 0.19}$     &$\textcolor{red}{4.046}$  & $\textcolor{red}{4.046}$ \\
 & $\textcolor{blue}{3.01\pm 0.18}$     & $\textcolor{blue}{3.08\pm 0.18}$    &$\textcolor{blue}{4.046}$  &$\textcolor{blue}{4.046}$  \\
 &$\textcolor{magenta}{2.98\pm 0.18}$     &$\textcolor{magenta}{2.98^{+0.19}_{-0.22}}$     &$\textcolor{magenta}{4.046}$  &$\textcolor{magenta}{4.046}$  \\
 &$\textcolor{darkgreen}{3.00^{+0.17}_{-0.18}}$     &$\textcolor{darkgreen}{2.97\pm 0.20}$     &$\textcolor{darkgreen}{4.046}$  &$\textcolor{darkgreen}{4.046}$  \\
                                                                                                  
{$\sum m_{\nu}$, $m_{\nu_s}   $} &$\textcolor{red}{0.12^{+0.01}_{-0.06},\;0  } $     &$\textcolor{red}{0.11^{+0.01}_{-0.05},\;0}$     &$\textcolor{red}{0.06,\;0.14^{+0.06}_{-0.11}}$  &$\textcolor{red}{0.06,\;0.17^{+0.09}_{-0.11}}$  \\
 &$\textcolor{blue}{0.135^{+0.020}_{-0.074},\;0}$     &$\textcolor{blue}{0.119^{+0.014}_{-0.059},\;0}   $     & $\textcolor{blue}{0.06,\;0.28^{+0.11}_{-0.13}}$  &$\textcolor{blue}{0.06,\;0.26^{+0.11}_{-0.13}}$  \\
 &$\textcolor{magenta}{0.135^{+0.022}_{-0.074},\;0}$     &$\textcolor{magenta}{0.235^{+0.072}_{-0.11},\;0}$     &$\textcolor{magenta}{0.06,\;0.37\pm 0.11 }$  &$\textcolor{magenta}{0.06,\; 0.59\pm 0.10}$  \\
 &$\textcolor{darkgreen}{0.13^{+0.03}_{-0.07},\;0}$     &$\textcolor{darkgreen}{0.23^{+0.08}_{-0.11},\;0}$     &$\textcolor{darkgreen}{0.06,\;0.37\pm 0.11 }$  &$\textcolor{darkgreen}{0.06,\; 0.60\pm 0.09}$  \\
                                                                                                 
{$\delta        $}  &$\textcolor{red}{-0.064^{+0.110}_{-0.083} }  $     &\textcolor{red}{0 }    &$\textcolor{red}{0.087\pm 0.068}$  &  \textcolor{red}{0}\\
 &$\textcolor{blue}{-0.143^{+0.075}_{-0.064}}   $     &\textcolor{blue}{0}     &$\textcolor{blue}{-0.077^{+0.077}_{-0.068}}$  &\textcolor{blue}{0}  \\
 &$\textcolor{magenta}{-0.157\pm 0.039 }$     &\textcolor{magenta}{0}     &$\textcolor{magenta}{-0.144\pm 0.040}$  &\textcolor{magenta}{0}  \\
 &$\textcolor{darkgreen}{-0.160\pm 0.040}$     &\textcolor{darkgreen}{0}     &$\textcolor{darkgreen}{-0.148\pm 0.041}$  &\textcolor{darkgreen}{0}  \\
\hline                                                                                                
{$H_0             $} &$\textcolor{red}{68.20\pm 1.40} $     & $\textcolor{red}{68.00\pm 1.20}$    &$\textcolor{red}{72.30\pm 1.00}$  & $\textcolor{red}{73.29\pm 0.75}$ \\
 &$\textcolor{blue}{68.9\pm 1.3 }$     &$\textcolor{blue}{68.0\pm 1.2 }$     &$\textcolor{blue}{73.6\pm 1.1}$  &$\textcolor{blue}{72.94\pm 0.83}$  \\
 &$\textcolor{magenta}{68.8\pm 1.3}$     &$\textcolor{magenta}{66.9\pm 1.2}$      &$\textcolor{magenta}{74.04\pm 0.99}$  &$\textcolor{magenta}{71.51^{+0.60}_{-0.74} }$  \\
 &$\textcolor{darkgreen}{69.00\pm 1.30}$     &$\textcolor{darkgreen}{66.90\pm 1.10}$      &$\textcolor{darkgreen}{74.06^{+0.90}_{-1.00}}$  &$\textcolor{darkgreen}{71.58\pm 0.62 }$  \\
                                                                                                 
{$\sigma_8 $} &$\textcolor{red}{0.791^{+0.035}_{-0.026}}$     &$\textcolor{red}{0.813\pm 0.013}$     &$\textcolor{red}{0.870^{+0.029}_{-0.025}}$  & $\textcolor{red}{0.841\pm 0.017}$ \\
 &$\textcolor{blue}{0.766^{+0.025}_{-0.022}}$     &$\textcolor{blue}{0.807\pm 0.013}$     &$\textcolor{blue}{0.799\pm 0.031}$  &$\textcolor{blue}{0.823\pm 0.020}$  \\
 &$\textcolor{magenta}{0.761\pm 0.010 }$     & $\textcolor{magenta}{0.770^{+0.012}_{-0.011}}$   &$\textcolor{magenta}{0.767\pm 0.011}$  & $\textcolor{magenta}{0.766^{+0.011}_{-0.013}}$ \\
 &$\textcolor{darkgreen}{0.761\pm 0.010}$     & $\textcolor{darkgreen}{0.769\pm 0.011}$   &$\textcolor{darkgreen}{0.767\pm 0.012}$  & $\textcolor{darkgreen}{0.765\pm 0.011}$ \\
                                                                                                  
%{$\Omega_{\rm m0}    $}  &$ \textcolor{red}{0.305\pm 0.010}$     &$\textcolor{red}{0.3081\pm 0.0080}$     &$\textcolor{red}{0.3012\pm 0.0094}$  &$\textcolor{red}{0.2946\pm 0.0081}$  \\
% &$\textcolor{blue}{0.2932\pm 0.0081}$     &$\textcolor{blue}{0.3007^{+0.0062}_{-0.0071}}$     &$\textcolor{blue}{0.3013\pm 0.0083}$  &$\textcolor{blue}{0.2998\pm 0.0077}$  \\
% &$\textcolor{magenta}{0.2977\pm 0.0080}$     & $\textcolor{magenta}{0.3128\pm 0.0083}$    &$\textcolor{magenta}{0.2922\pm 0.0092}$  & $\textcolor{magenta}{0.3110\pm 0.0074}$ \\
% &$\textcolor{darkgreen}{0.2911\pm 0.0066}$ &$\textcolor{darkgreen}{0.3038\pm 0.0074}$   & $\textcolor{darkgreen}{0.298\pm 0.008}$         &   $\textcolor{darkgreen}{0.309\pm 0.007}$ \\                
\hline
{$\chi^2_{\rm min}/2$} &$\textcolor{red}{6821.18}$     &$\textcolor{red}{6821.63}$     &$\textcolor{red}{6833.61}$  &$\textcolor{red}{6833.82}$  \\
 &$\textcolor{blue}{6822.20}$     &$\textcolor{blue}{6825.12}$     & $\textcolor{blue}{6834.72}$  &$\textcolor{blue}{6836.03}$  \\
 &$\textcolor{magenta}{6822.64}$     & $\textcolor{magenta}{6831.06}$    &$\textcolor{magenta}{6835.72}$  & $\textcolor{magenta}{6841.63}$ \\
 &$\textcolor{darkgreen}{6821.98}$     & $\textcolor{darkgreen}{6831.07}$    &$\textcolor{darkgreen}{6836.14}$  & $\textcolor{darkgreen}{6841.75}$ \\
\hline
\end{tabular}
\end{center}
\label{tab1}
\end{table*}
%#################################################################################################################################

\subsection{Analysis of constraints on $\nu$IVCDM model}

The second column of Table \ref{tab1} shows the constraints (mean values with 1$\sigma$ errors) on the free parameters, 
and two derived parameters $H_0$ and $\sigma_8$ of the $\nu$IVCDM model
from the four different combinations of data sets as mentioned in the caption of the table. 
In contrast, the third column of Table \ref{tab1} shows the results for the $\nu\Lambda$CDM model.

Figure \ref{fig1} shows the one-dimensional marginalized distribution and 68\% CL, 95\% CL regions
% for some selected parameters $N_{\rm eff}$, $H_0$, $\sigma_8$ and $\delta$ pertaining to the $\nu$IVCDM model from the four combinations 
for some selected parameters of the $\nu$IVCDM model from the four combinations 
of data sets as shown in the legend of the figure. We do not find evidence of interaction in the dark sector from the minimal CMB + BAO + JLA data set, 
since the constraints on $\delta$ are closed to be null even at 68\% CL.
On the other hand, the inclusion of GC data to the CMB + JLA + BAO data can yield non-zero $\delta$ up to $99\%$ CL. 
Here, we considered three different combinations by including CFHTLenS and SZ data separately and then jointly to our minimum data set. The inclusion of CFHTLenS alone gives $\delta < 0$ at 95\% CL.
We notice no significant change in the constraints yielded by the inclusion of SZ alone and CFHTLenS + SZ. We find $- 0.060  \leq \delta \leq - 0.260$ at $99\%$ CL from CMB + JLA + BAO + CFHTLenS + SZ data.
This shows a strong evidence of interaction in the dark sector within the framework of the $\nu$IVCDM model under 
these considerations. Also, we notice from our results that $\delta < 0$, which in turn implies that the DM 
decays into vacuum energy. Similar results are reported in \cite{Salvatelli} for the late-time interaction 
in the dark sector, but from another observational perspective where the authors considered different redshift bins 
for the constraints.

Further, we notice a significant correlation of the coupling parameter $\delta$ with $H_0$ and $\sigma_8$.
Larger values of $H_0$ and smaller values of $\sigma_8$ correspond to a correlation into direction of a non-zero and
negative values of $\delta$. We have observed that this combined effect can play an important 
role in a possible interaction in the dark sector. 
%From the constraints on $H_0$, we note that its best fit value lies in the vicinity of 69 km/s/Mpc. 
From the constraints, we can observe that the parameter $\delta$
is pushed in the non-zero negative range by GC data, and hence the correlation of $\delta$ with other parameters is induced by GC data.
We also note a positive correlation between $N_{\rm eff}$ and $H_0$ (very well known in the literature), 
indicating that larger values of $N_{\rm eff}$ would correspond to larger values of $H_0$. 
Therefore, extra species of neutrinos would result into larger values of $H_0$. Once
$H_0$ is correlated with $\delta$, larger values of $H_0$ can also push  
$\delta$ into negative range with further smaller values. The larget data set CMB + BAO + JLA + CFHTLenS + SZ under consideration yields
$N_{\rm eff} < 3.45 $ and $\sum m_{\nu} < 0.31 $ eV at 99\% CL.

\subsection{ Analysis of constraints on $\nu_s$IVCDM model}

As a generalization of the $\nu$IVCDM model, it is natural to investigate the cosmological consequences with 
sterile neutrinos, i.e., we study the $\nu_s$IVCDM model, the (3+1) model as described in section \ref{model}. 
Since $N_{\rm eff} < 3.45 $ and $\delta \neq 0$ at $99\%$ CL from CMB + BAO + JLA + CFHTLenS + SZ data in the $\nu$IVCDM model, 
it would be interesting to investigate the correlation $N_{\rm eff}$, and the mass $m_{\nu_s}$ of sterile neutrino with $\delta$, 
and the other baseline parameters of the $\nu_s$IVCDM model. The third column of Table \ref{tab1} carry the constraints 
(mean values with 68\% errors) on some free parameters of the $\nu_s$IVCDM model from four different combinations of data sets as 
described in the caption of the table. Constraints for $\nu_s\Lambda$CDM model are also shown. 

Figure \ref{fig2} shows the one-dimensional marginalized distribution, and 68\% and 95\% CL regions 
for some selected parameters of the $\nu_s$IVCDM model from the four combinations of data sets as shown in the legend of the figure.
We see that the data sets CMB + JLA + BAO again do not provide any significant evidence of interaction in the dark sector. 
On the other hand, the inclusion of CFHTLenS data does not provide significant deviations on the coupling parameter as
$\delta \sim 0$ at 68\% CL. But the inclusion of SZ data alone and SZ + CFHTLenS data
yield $\delta < 0$ at 99\% CL. More precisely, we note the coupling parameter 
in the range $- 0.04 \leq \delta \leq - 0.25$ at $99\%$ CL. This shows again a strong evidence of interaction in the dark sector 
when the GC data in taken into account. Further, we observe a correlation between $\delta$ and $m_{\nu_s}$ in case of the CMB + JLA + BAO data set
with and without GC data. The inclusion of GC data brings a correlation between the two parameters. 
The parameter $\delta$ exhibits correlation with both $H_0$ and $\sigma_8$.
Thus, the larger values of $H_0$ and smaller values of $\sigma_8$ would push the parameter $\delta$ 
(with negative values) away from 0. Likewise, correlations among the parameters $m_{\nu_s}$, $H_0$, and $\sigma_8$ can be seen clearly 
from the Figure \ref{fig2}. From the joint analysis using all the data sets under consideration, we find $m_{\nu_s} = 0.37^{+0.11}_{-0.11}$ eV at 68\% CL.

It is well known that the parameters describing neutrinos can exhibit degeneracy 
with the other cosmological parameters. In \cite{Li}, the authors show that there is a strong degeneracy between the spectral index 
($n_s$) and $N_{\rm eff}$. As discussed in \cite{Planck_inflation_2015,Planck_inflation_2013}, the $\Lambda$CDM scenario with $n_s = 1$ 
and $\Delta N_{\rm eff} \approx 1$ can fit the Planck CMB data. In \cite{Verde}, the authors considered 
$N_{\rm eff}$ + $\Lambda$CDM (with $n_s = 1$), and showed that the model can perfectly fit CMB data. 
Therefore, we can expect a correlation between the parameters $N_{\rm eff}$, $n_s$ and $H_0$.
The unusual high constraints on $n_s$ observed in $\nu_s$IVCDM model are surely due to the correlation between $N_{\rm eff}$ and $n_s$, 
which is related to the diffusion damping caused by $N_{\rm eff}$ at high multipoles,
that requires $n_s \sim 1$ to compensate the suppression. On the other hand, the additional free parameters representing physics 
beyond the $\Lambda$CDM model can also correlate with these parameters. Thus, the constraints on $n_s$ also could be the outcome of
that correlation. A detailed analysis of an extended parametric space for cosmological models besides $\Lambda$CDM scenario, 
including effects on inflation parameters, will be presented in a forthcoming paper \cite{eu}.

\subsection{Analysis of the tensions on $H_0$ and $\sigma_8$}

\begin{figure*}
\includegraphics[width=7.0cm]{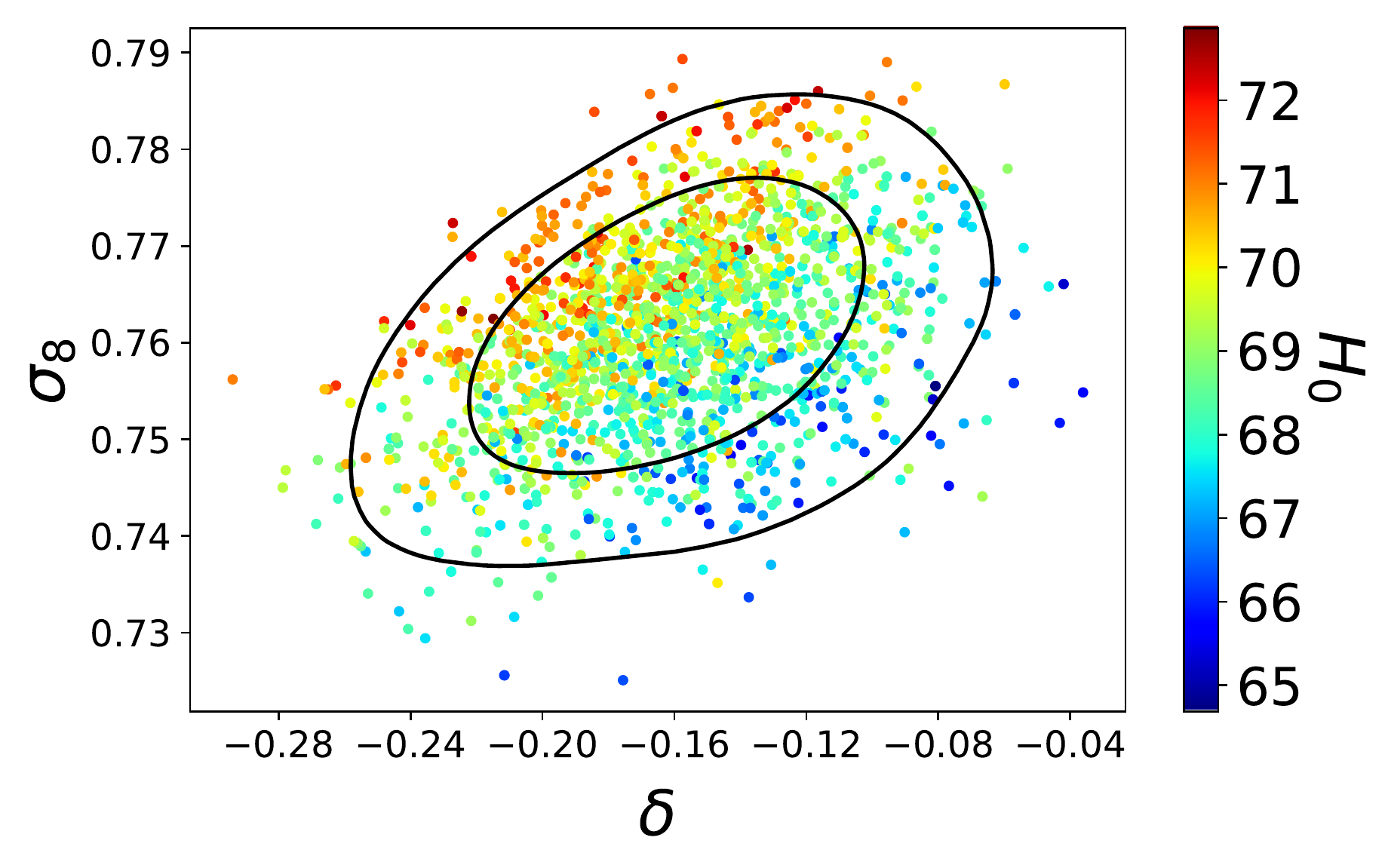}
\includegraphics[width=7.0cm]{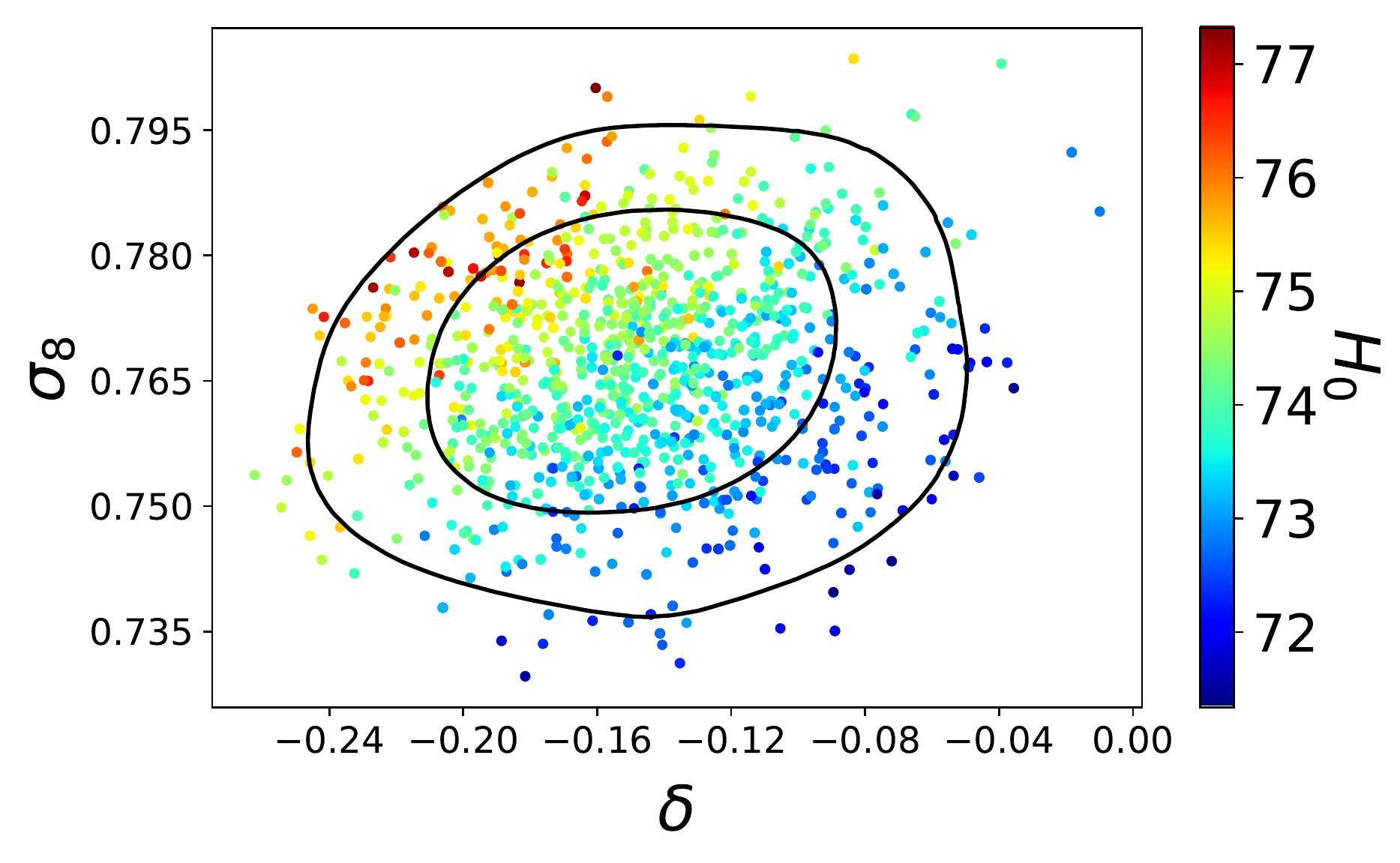}
\caption{\label{fig3} {\it{Constraints on $\sigma_8 - \delta$ parametric space coloured by $H_0$ values 
for the $\nu$IVCDM (left panel) and $\nu_s$IVCDM (right panel) models from CMB + BAO + JLA + CFHTLenS + SZ data.}}}
\end{figure*}

Assuming standard $\Lambda$CDM baseline, the Planck collaboration \cite{Planck2015}
measured $H_0 = 67.27 \pm 0.66$ km s${}^{-1}$ Mpc${}^{-1}$ that is about
two standard deviations away from the locally  measured value $H_0=73.24 \pm 1.74$ km s${}^{-1}$ Mpc${}^{-1}$ 
reported in \cite{Riess}. Results from Planck collaboration have also revealed around 2$\sigma$ level tension between 
CMB temperature and the Sunyaev-Zel'dovich cluster abundances measurements on $\sigma_8$ parameter given by $\sigma_8 = 0.831 \pm 0.013$ \cite{Planck2015} 
and $\sigma_8 =0.75 \pm 0.03 $  \cite{S8_planck}, respectively. Thus, the galaxy cluster measurements prefer lower values of $\sigma_8$.
In \cite{Pourtsidou}, it has been argued that interacting dark sector models can 
reconcile the present tension on $\sigma_8$. In the present study, we have observed strong correlation of the coupling parameter 
$\delta$ with $H_0$ and $\sigma_8$, in both $\nu$IVCDM and $\nu_s$IVCDM models, such that the larger departure of negative $\delta$ values
from 0 leads to larger values of $H_0$ and smaller values of $\sigma_8$. Also, we can see the degeneracy between $\delta$, $H_0$, and $\sigma_8$ 
with clarity in Figure \ref{fig3}, where the parametric space $\sigma_8 - \delta$ coloured by $H_0$ values is shown 
for the $\nu$IVCDM (left panel) and $\nu_s$IVCDM (right panel) models. Having a note of these observations, 
we now proceed to examine whether these models can alleviate the current tensions on $H_0$ and $\sigma_8$.

\begin{figure*}
\includegraphics[width=7.0cm]{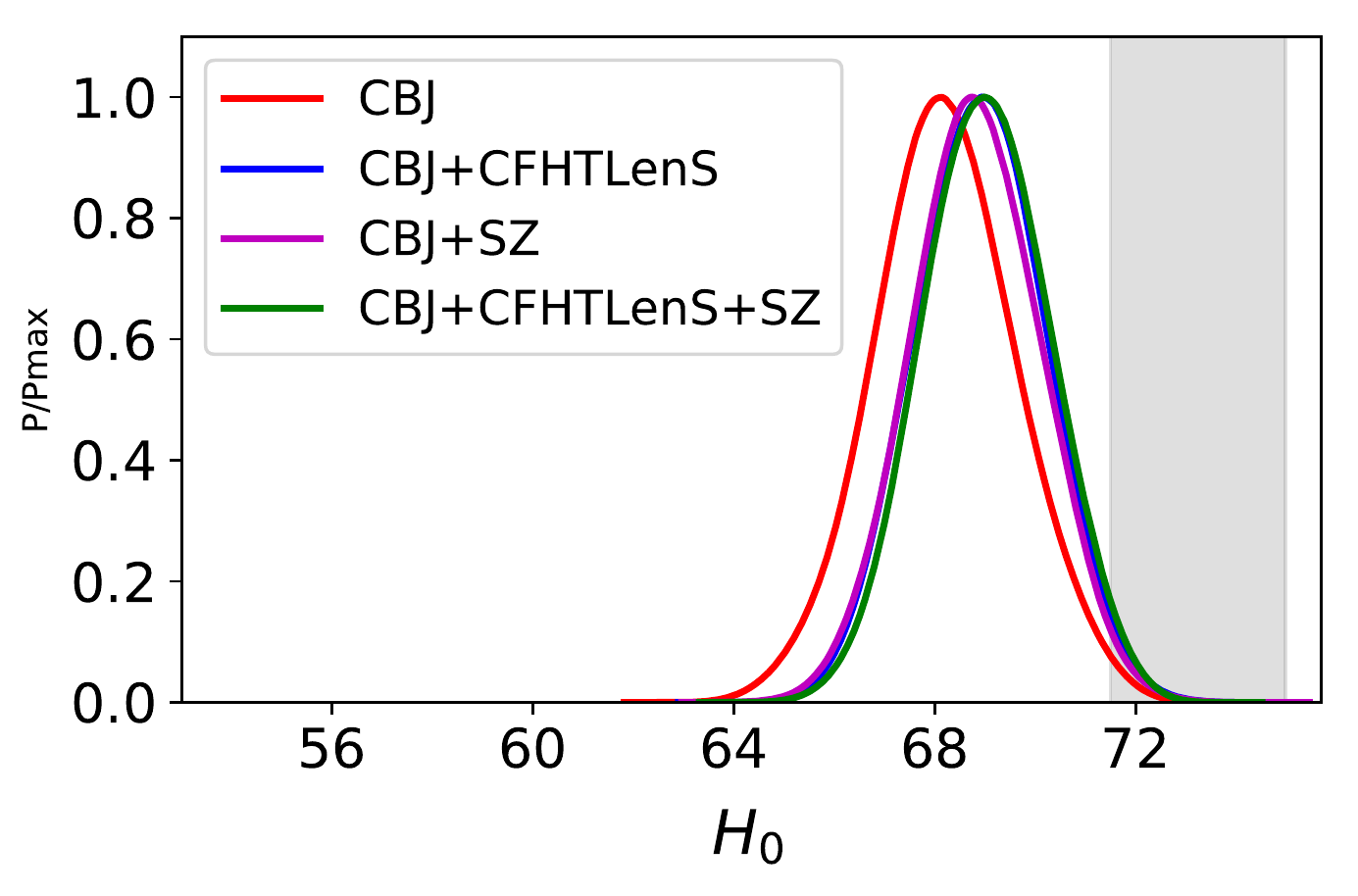}
\includegraphics[width=7.0cm]{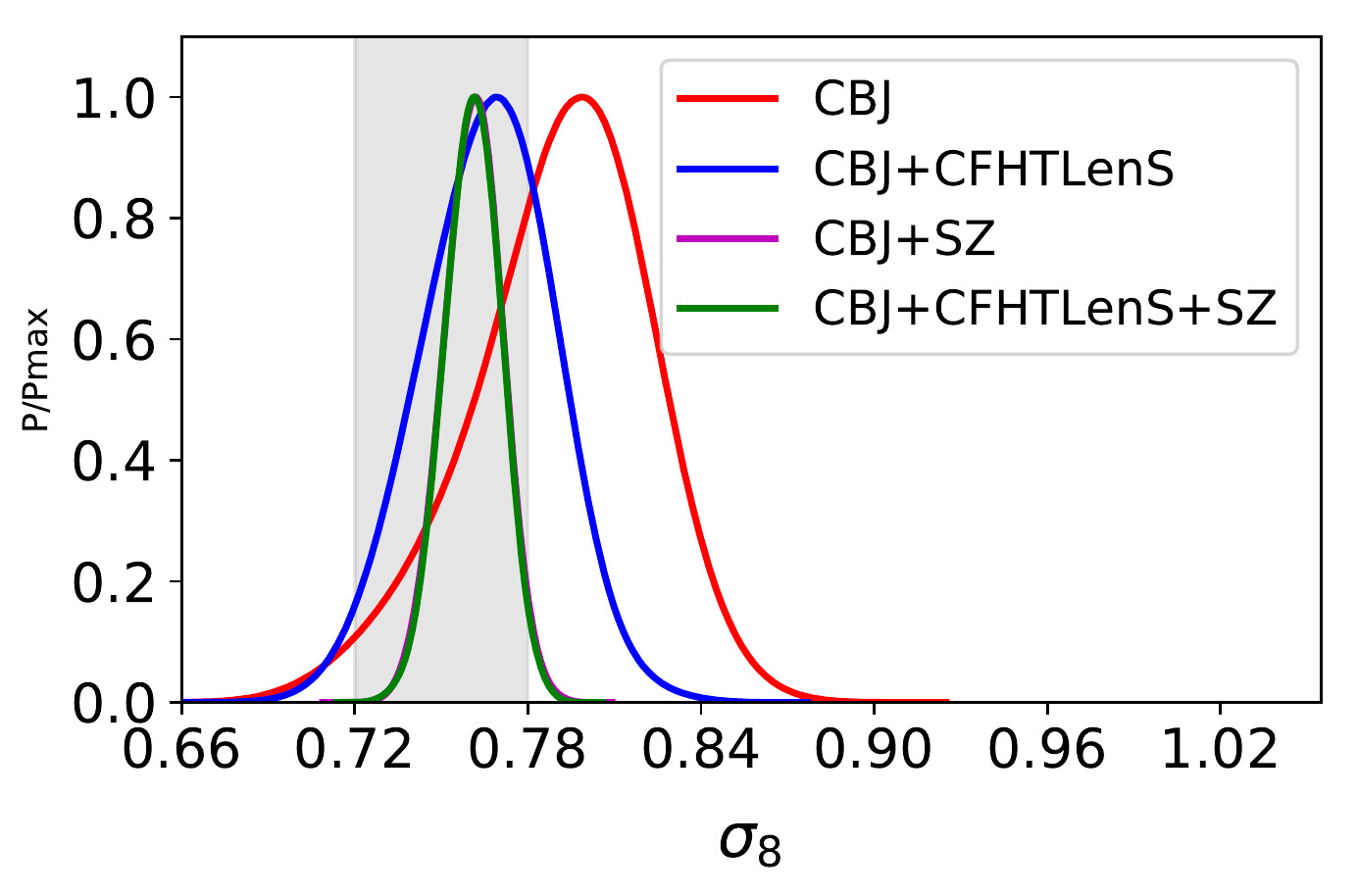}
\includegraphics[width=7.0cm]{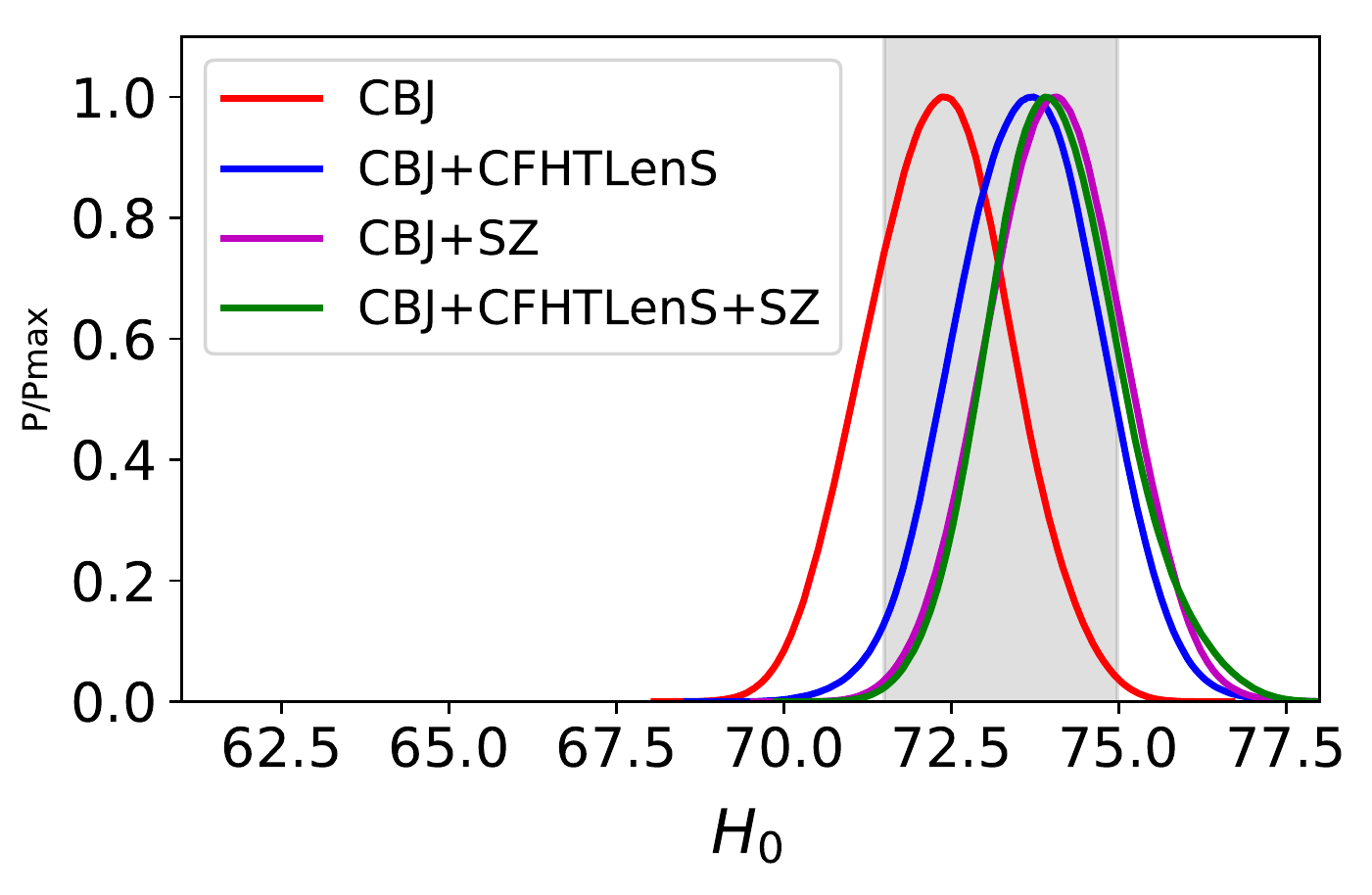}
\includegraphics[width=7.0cm]{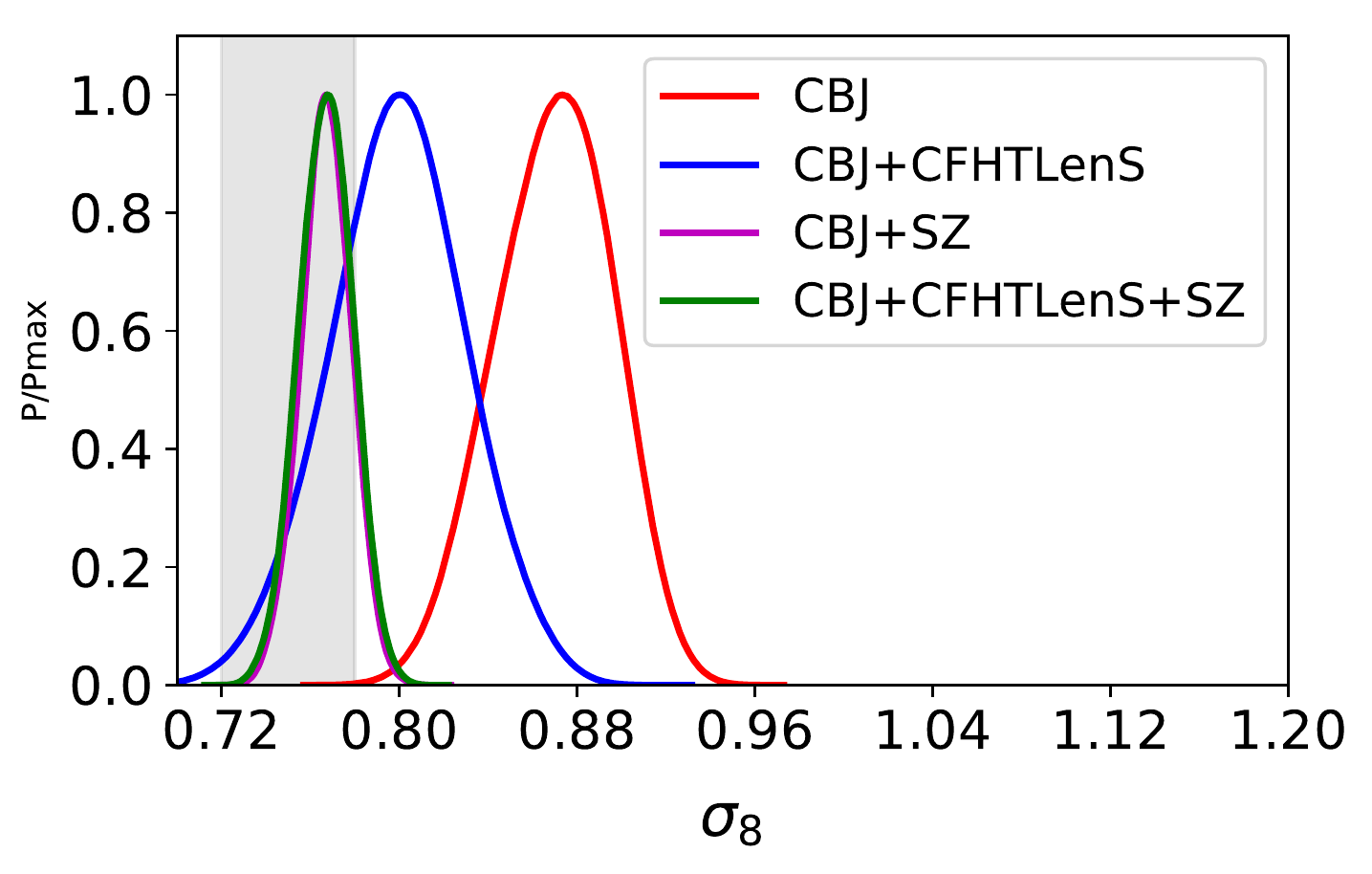}
\caption{\label{fig4} {\it{1D marginalized probability distributions of $H_0$ and $\sigma_8$ for $\nu$IVCDM (upper panel) 
and $\nu_s$IVCDM (lower panel) models. The vertical gray band in left panels corresponds to $H_0 = 73.24 \pm 1.74$ km s${}^{-1}$Mpc \cite{Riess} 
while in right panels it corresponds to $\sigma_8 = 0.75 \pm 0.03$ \cite{S8_planck}. CBJ stands for CMB + BAO + JLA.}}}
\end{figure*}

Figure \ref{fig4} shows 1D marginalized probability distributions of $H_0$ and $\sigma_8$ for $\nu$IVCDM (upper panel) and $\nu_s$IVCDM 
(lower panel) models. We notice that the tension on $H_0$ can not be reconciled marginally in the $\nu$IVCDM model. 
The plot in upper right panel shows that tension on $\sigma_8$ is reconciled perfectly in the $\nu$IVCDM model when the GC data is taken 
into account for the constraints. See the constraints on $H_0$ and $\sigma_8$ in second column of Table \ref{tab1}. 
It is important to mention that we are not taking any prior on $H_0$ in our analysis.
From the plot in lower left panel, we notice that the tension on $H_0$ is reconciled in $\nu_s$IVCDM model when constrained with any of the 
four data combinations under consideration. This is due to the positive correlation between $N_{\rm eff}$ and $H_0$.
On the other hand, the tension on $\sigma_8$ is reconciled in $\nu_s$IVCDM model when it is constrained by including GC data from SZ alone 
or SZ + CFHTLenS (see the plot in the lower right panel of Figure \ref{fig4}). 
Thus, the $\nu_s$IVCDM model can alleviate the tensions on both the parameters $H_0$ and $\sigma_8$,  
with the coupling parameter $\delta \neq 0$ at 99\% CL from the full data set under consideration.
A detailed and clear exposition of all the values of these parameters in $\nu_s$IVCDM model, from all the four data combinations,
can be seen in the third column of table \ref{tab1}.
\begin{table*}[!htbp]
\caption{\label{tab2} Summary of the $\Delta \rm AIC$ values and its interpretation. 
          Here, $\Delta \rm AIC_{\nu} = AIC(\nu \Lambda CDM) - AIC(\nu IVCDM)$ and $\Delta \rm AIC_{\nu_s} = AIC(\nu_s \Lambda CDM) - AIC(\nu_s IVCDM)$.}
     \begin{center}
\begin{tabular} { l  l l}
\hline
 Data combination  &    $\Delta \rm AIC_{\nu}$    & $\Delta\rm AIC_{\nu_s}$  \\
\hline
{CMB+BAO+JLA}              &  $-1.10$ (no evidence)           & $-1.58$ (no evidence)    \\
{CMB+BAO+JLA+CFHTLenS}     &  3.84  (positive evidence)     &  0.62  (no evidence)    \\                                                                                              
{CMB+BAO+JLA+SZ}           &  14.84 (very strong evidence)  &  9.82  (strong evidence)   \\
{CMB+BAO+JLA+CFHTLenS+SZ}  &  16.18 (very strong evidence)  &  9.21  (strong evidence)  \\ \hline
\end{tabular}
\end{center}
\label{tab2}
\end{table*}
It is usual in the literature to reconcile both the tensions by assuming 
an extended parametric space of $\Lambda$CDM model. 
The authors in \cite{T4} argue that the tension must be resolved by considering systematics effects on important data, 
or by new physics beyond the introduction of massive (active or sterile) neutrinos. 
Here we have demonstrated the reconciliation of both the tensions, 
while observing the signs of a new physics, viz., the coupling parameter of the dark sector is non-zero up to $99\%$ CL.

%\section{Model comparison}

Finally, we close the observational analysis section by comparing the statistical fit obtained 
using the standard information criteria via Akaike Information Criterion (AIC) \cite{Akaike,Burnham}

\begin{align}
\label{AIC}
\rm AIC = - 2 \ln L + 2d = \chi^2_{min} + 2d,
\end{align}
where $L = \exp(-\chi^2_{min}/2)$ is the maximum likelihood function and $d$ is the number of model parameters.

We need a reference model with respect to which the comparisons will be performed. In this regard, the obvious choice is the $\Lambda$CDM cosmology with and without sterile neutrinos in comparison with IVCDM model with and without sterile neutrinos.

The thumb rule of AIC reads as follows: If $\Delta\rm  AIC \leq 2$, 
then the models are compatible, i.e., statistically indistinguishable from each other. The range $ 2 \leq  \Delta\rm  AIC \leq 6$ 
tells a positive evidence, $ 6 \leq  \Delta\rm  AIC \leq 10$ implies a strong evidence and $\Delta\rm  AIC \geq 10$
infers a very strong evidence. 

Table \ref{tab2} summarizes the comparison between the models for each analysis.
Without loss of generalization, these results are also valid with other standard criteria in the literature
evaluating directly the values of $\chi^2_{min}$. From Table \ref{tab2}, we notice that the inclusion of SZ data to the other data sets in the fit leads to strong evidence. It is in line with the results in Table \ref{tab1} where one may observe that the $\chi^2_{min}$ values increase significantly with the inclusion of SZ data. From Table \ref{tab1} one may further notice that the two models with sterile neutrino fit with larger $\chi^2_{min}$ values in comparison to the other two models whereas the interacting vacuum energy models fit with smaller $\chi^2_{min}$ values in comparison to the non-interacting ones. It is well known that a model with $\Delta N_{\rm eff} = 1$ can be
disfavored up to 3$\sigma$ CL when confronted with the observational data. For instance, Planck team \cite{CMB} using CMB + BAO data excluded the possibility $N_{\rm eff} = 4$ at 99\% CL, in comparison to the case where $N_{\rm eff}$ is taken as a free parameter.
Thus, it is reasonable to expect that the value of  $\chi^2_{min}$ increases
in our analyzes too where we fix $\Delta N_{\rm eff} = 1$. In Table \ref{tab1},  we notice that the joint analysis yields, $\Delta \chi^2 = 10.68$ (for $\Lambda$CDM model) and $\Delta \chi^2 = 14.16$ (for the interacting vacuum energy model), when compared the 3 + 1 neutrino model to the case where $N_{\rm eff}$ is free parameter. On the other hand, analyzing the joint analysis fit only within the 3 + 1 neutrino model scenario, we can notice that the vacuum decay model improves the fit by $\Delta \chi^2 = 5.61$ when compared with the $\Lambda$CDM model.

\section{Conclusions}

We have shown how an extended parametric space with massive neutrinos can influence the observational constraints 
on a coupling between DM and DE, where DE is characterized by a cosmological constant, i.e., an interacting vacuum energy scenario.
Considering two neutrino scenarios via  $\nu$IVCDM and  $\nu_s$IVCDM models, we have found that the inclusion of GC data 
to CMB + BAO + JLA data considerably improves the observational constraints in favor of an interaction 
in the dark sector. In particular, both the models yield the coupling parameter $\delta < 0$ 
at 99\% CL when constrained by including the GC data. Thus, we conclude that the GC data enforces almost a certain 
interaction in the dark sector where the DM decays into vacuum energy, within the framework of each of the two models under consideration. 
We have demonstrated how the coupling parameter $\delta$ correlates with other parameters, and how the current tensions on the parameters 
$H_0$ and $\sigma_8$ can be reconciled in the two models.

Recent anomalies in neutrino experiments can be explained by neutrino oscillations 
if the standard three-neutrino mixing is extended with the addition of light
sterile neutrinos, which can give us important information on the new physics beyond
the standard model. In the present study, we have studied the so-called (3 + 1) neutrinos model, i.e., the $\nu_s$IVCDM model, and 
also we have found an evidence for a coupling between DM and DE at 99\% CL.
If future neutrino experiments confirm the existence of light sterile neutrinos, 
this may lead to a possible evidence for a non-minimal coupling between the dark components 
on cosmological scales. 

In the present work, we have considered GC data from $\sigma_8 - \Omega_m$ plane ($S_8$ data is usually used in the literature) and 
we noticed a strong correlation with the coupling parameter between DM and DE. It is important to mention that this data set
consists of a useful parameterization, but a detailed analysis using the full likelihood (direct information from the 
power spectrum) can lead to more accurate and useful results. In general, the GC data depends on non-linear physics. 
Further, it is also well known that massive neutrinos play an important role on small scale (non-linear scale). Thus, 
investigation of interacting DE models within this perspective may be worthwhile.

Future and more accurate measurements from important cosmological sources, in particular
large scale structure (including non-linear scale) and CMB, and of neutrino properties including 
mass hierarchy, effective number of species, as well as the sum of the 
active neutrino masses, can shed light on the possible interaction in the 
dark sector.

\section*{Acknowledgments}

%\textbf{The authors thank the anonymous referees for his/her detailed comments and suggestions which improved the work signicantly}. 
The authors are grateful to S. Gariazzo, E. D. Valentino, J. Alcaniz, and S. Pan for useful comments.
S.K. gratefully acknowledges the support from SERB-DST project No. EMR/2016/000258.

\end{document}